\documentclass[10pt,letterpaper]{article}
\usepackage{opex3}

\begin{document}

%%%%%%%%%%%%%%%%%% title page information %%%%%%%%%%%%%%%%%%
\title{Observation of Goos-H$\rm \ddot{\textbf{a}}$nchen shifts in metallic reflection}

\author{M. Merano, A. Aiello, G. W. 't Hooft, M. P. van Exter, E. R. Eliel

and J. P. Woerdman}

\address{Huygens Laboratory, Leiden University, P.O. Box 9504, 2300 RA Leiden, The Netherlands}

\email{merano@molphys.leidenuniv.nl} %% email address is required

% \homepage{http:...} %% author's URL, if desired

%%%%%%%%%%%%%%%%%%% abstract and OCIS codes %%%%%%%%%%%%%%%%
%% [use \begin{abstract*}...\end{abstract*} if exempt from copyright]

\begin{abstract*}
We report the first observation of the Goos-H$\rm \ddot{\textbf{a}}$nchen shift of a light beam incident on a bare metal surface. This phenomenon is particularly interesting because the Goos-H$\rm \ddot{\textbf{a}}$nchen shift for $p$ polarized light in metals is negative and much bigger than the positive shift for $s$ polarized light. The experimental result for the measured shifts as a function of the angle of incidence is in excellent agreement with theoretical predictions. In an energy-flux interpretation, our measurement shows the existence of a backward energy flow at the bare metal surface when this is excited by a $p$ polarized beam of light.
\end{abstract*}

 % REPLACE WITH CORRECT OCIS CODES FOR YOUR ARTICLE

%%%%%%%%%%%%%%%%%%%%%%% References %%%%%%%%%%%%%%%%%%%%%%%%%

%%%%%%%%%%%%%%%%%%%%%%%%%%  body  %%%%%%%%%%%%%%%%%%%%%%%%%%
\section{Introduction}
The Goos-H$\rm \ddot{a}nchen$ (GH) shift \cite{Goos47} is the displacement, with respect to geometrical reflection, of an $s$ or $p$ polarized light beam reflected by a medium with a complex and angle-dependent reflection coefficient (Fig. 1); it is essentially a diffractive correction on geometrical optics. The incident beam can be considered as the superposition of plane waves that upon reflection experience different phase shifts. By a stationary phase method, originally due to Artmann \cite{Artmann48}, it is possible to show that the sum of these slightly phase-shifted plane waves results in a reflected beam that is laterally displaced in the plane of incidence. If $\delta(\theta)$ is the phase of the complex reflection coefficient it can be shown that the GH shift ($D$) is given by
\begin{equation}
 D=-\frac{\lambda}{2\pi}\frac{d\delta(\theta)}{d\theta}
\end{equation}
where $\lambda$ is the wavelength of light and $\theta$ is the angle of incidence. If the incoming beam is at some definite polarization state, the reflected light will consist of two beams, one displaced by $D_{s}$ (the $s$ polarized component) and one by $D_{p}$ (the $p$ polarized component) \cite{Chiu74, Gragg88}.

\begin{figure}[h]
\centering\includegraphics[scale=0.5]{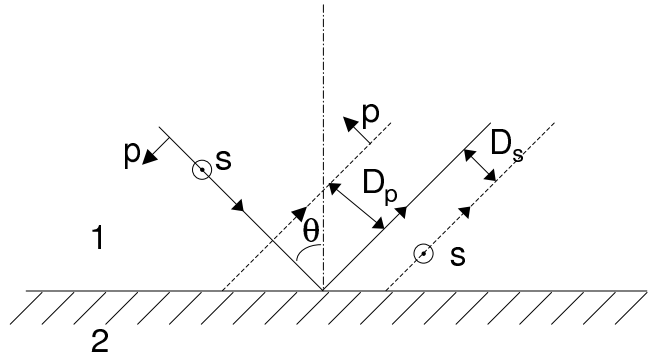}% Here is how to import EPS art
\caption{\label{fig:epsart}Geometry indicating the GH shift, defined as D. A beam of light with a finite transverse extent is incident from vacuum (medium 1) on a metal surface (medium 2). If the beam is $s$ polarized, the displacement of the reflected beam (dotted line) with respect to the geometrical reflection (continuous line) is positive. If the beam is $p$ polarized, the displacement is negative.}
\end{figure}

Goos and H$\rm \ddot{a}nchen$ \cite{Goos47} were the first to experimentally demonstrate this shift for the case of total internal reflection (TIR) from the surface of an insulator with dielectric constant $\epsilon$. In their experiment they employed multiple reflections in a glass slab, in order to amplify the small displacement of the beam on a single reflection. More recently, other groups mounted original set ups to measure the GH shift; they were successful in measuring the dependence of the GH shift on the angle of incidence in a single-reflection experiment \cite{Bretenaker92, Gilles02}.

The GH shift has also been interpreted as a proof of the existence of a flow of energy parallel to the surface inside the less dense medium \cite{Lai00, Renard64, Lotsch68}. The existence of such a flux of energy has in fact been debated from the early days of the GH effect. An experiment that shed more light on this aspect of the GH shift was performed by Rhodes and Carniglia \cite{Carniglia77}. By an interferometric technique, they proved evidence of the shift near grazing incidence in a TIR configuration; this case was controversial at the time. Their experiment falsified some \cite{Renard64, Lotsch68}, but not all, of the theories that explained the GH effect in term of energy-fluxes. More recently, new calculations of energy-flux patterns in the GH effect were developed \cite{Lai00}; they are compatible with the theory of Artmann.

The fact that in the paradigmatic case of TIR the GH effect is positive raised interest in systems where the GH shift is negative. Theoretical evidence of negative GH shifts in periodic structures has been reported \cite{Tamir71, Anicin78}. Recently, theoretical works, have predicted negative GH shifts also in photonics crystals and left-handed materials \cite{He06, Wang05}. Experimental evidence of negative GH shifts has been obtained in measurements of lateral displacement of an optical beam enhanced by surface plasmon excitation \cite{Hesselink04, Bretenaker01}. However, in these cases the GH effect is essentially altered because a propagating surface energy flux is artificially created.

Here we report, as a paradigmatic example of a negative GH effect, the observation of the GH shift in conventional \emph{metallic reflection}.

\section{Theoretical predictions}

\begin{figure}
\centering\includegraphics[scale=0.45]{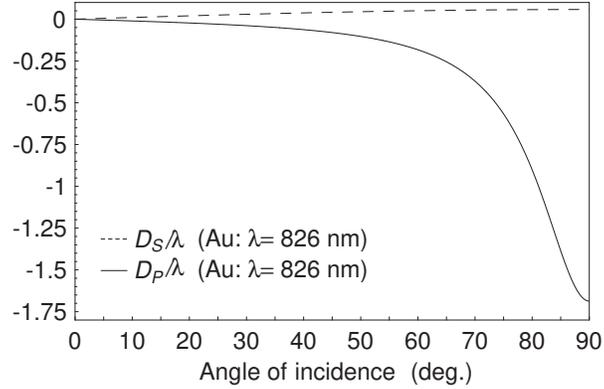}% Here is how to import EPS art
\caption{\label{fig:epsart} Curves representing the theoretical GH shift (normalized to the wavelength of light) for reflection by an Au surface. We used the experimental optical constants of Au at 826 nm \cite{Palik}. It is important to note that while $D_{p}$ is negative, $D_{s}$ is positive and that $|D_{p}|\gg|D_{s}|$.}
\end{figure}

Wolter \cite{Wolter50, Lotsch70} was the first to consider the GH effect theoretically when the second medium is a metal (Fig. 1). Theoretical curves for the GH shift of an $s$ or $p$ polarized collimated beam at a vacuum-metal interface were later presented by Wild and Giles \cite{Giles82} and by Leung \emph{et al.} \cite{Leung07}. In particular, it has been emphasized by Leung \emph{et al.} \cite{Leung07} that metallic reflection has the advantage that the reflected beam at large angles of incidence is hardly suppressed by the (pseudo) Brewster effect. This is contrary to the case of weakly absorbing media (e.g. semiconductors at visible wavelengths) where it is the suppression of the reflected beam near the Brewster angle that hinders the observation of a large GH shift \cite{Giles82, Lai07}.

The phase of the complex reflection coefficients for the $s$ and $p$ polarizations is \cite{Aiello}:
\begin{equation}
\delta_{s}(\theta)=\Im m\left(\ln\left[\frac{n_{1}\cos(\theta)-(\hat{n}_{2}^{2}-n_{1}^{2}\sin^{2}(\theta))^{1/2}}{n_{1}\cos(\theta)+(\hat{n}_{2}^{2}-n_{1}^{2}\sin^{2}(\theta))^{1/2}}\right]\right),
\end{equation}
\begin{equation}
\delta_{p}(\theta)=\Im m\left(\ln\left[\frac{\hat{n}_{2}^{2}\cos(\theta)-n_{1}(\hat{n}_{2}^{2}-n_{1}^{2}\sin^{2}(\theta))^{1/2}}{\hat{n}_{2}^{2}\cos(\theta)+n_{1}(\hat{n}_{2}^{2}-n_{1}^{2}\sin^{2}(\theta))^{1/2}}\right]\right),
\end{equation}
respectively, where $\Im m$ indicates the imaginary part of a complex number, $n_{1}$ is the real index of refraction of the incident medium (air in our case), and the metal has a complex refractive index $\hat{n}_{2}=n_{2}+ ik$, where $k$ is the extinction coefficient. Using Eqs. (1), (2) and (3) we can compute the expected curves of the GH shift as a function of the angle of incidence. In Fig. 2 we show the theoretical results for the case that light at a wavelengh of 826 nm is incident on a Au mirror; we use as experimental value for the complex index of refraction of Au, $\hat{n}_{2}=0.188+ i 5.39$ \cite{Palik}. It is important to note the striking negative shift of the $p$ polarized beam, and also the fact that $|D_{p}|\gg|D_{s}|$.

\section{Experiment}

Our experimental set up is sketched in Fig. 3. A collimated gaussian beam at a wavelength of 826 nm is incident at a given angle on a Au mirror. With a quadrant detector (New Focus, model 2901), denoted as QD, we measure the displacement of the laser beam in the plane of incidence when the polarization of the beam is switched from $p$ to $s$.  The laser source is a temperature controlled near-infrared single-mode fiber-pigtailed laser diode (Thorlabs, model LPS-830-FC) that provides a cw beam at a wavelength of 826 nm. A microscope objective collimates the beam that leaves the exit facet of the fiber. The $1 / e^{2}$ intensity radius $R$ of the collimated beam after the microscope objective is 1.62 mm. An inverted beam expander is used to reduce the beam radius to 860 $\rm \mu$m. This choice is dictated by the fact that we want to measure GH shifts at angles up to 87$^{\circ}$ (grazing incidence) and by the dimension of the QD, that has a square active region of 3x3 mm. The beam is $p$ polarized by means of a Glan polarizing prism. Subsequently its polarization is switched between $p$ and $s$ at a frequency of 2.5 Hz with a nematic liquid-crystal variable retarder (LCVR) driven by a square-wave voltage. After reflection upon the Au mirror, the QD signal is fed into a lock-in amplifier in order to detect the beam displacements. The QD is mounted on linear translation stages (horizontal and vertical movements) that allow for optimal centering on the reflected beam. A nanovoltmeter (Keithley 181) is used to check optimal centering.

\begin{figure}
\centering\includegraphics[scale=0.45]{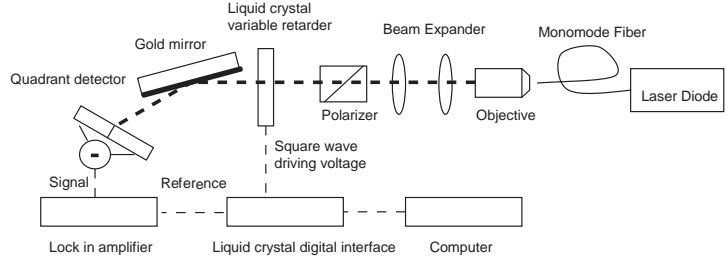}% Here is how to import EPS art
\caption{\label{fig:epsart} Schematic drawing of the experimental set up.}
\end{figure}

The QD consists of four distinct photodiodes, isolated from each other by a small gap. This gap is 100 $\rm \mu m$, i.e. much smaller than the diameter of the incident beam. The beam of light hits the detector orthogonally; the position of the centroid, with respect to the center of the detector, can be determined by analyzing the detector signal from all four quadrants. The QD provides three output channels: the SUM channel, the X channel and the Y channel. The SUM channel provides a voltage proportional to the beam intensity. The X channel provides a signal ($\chi$) that is equal to $\chi =C I d$, where C is a numerical factor that depends on the beam shape and on the responsivity of the diode, $I$ is the beam intensity, and $d$ is the displacement in the plane of incidence, of the centroid of the beam from the center of the QD ($d\ll R$). The Y channel instead measures displacements in the plane orthogonal to the plane of incidence. The position calibration of the QD can be performed indirectly by making the assumption that the beam is Gaussian and by measuring the beam diameter, the beam power (4.22 mW after the LCVR), and the voltages at the outputs of the QD. We made also a direct calibration procedure by inserting in the beam a 100 $\rm \mu$m thick plane parallel window that we rotated over a small angle. The two procedures gave the same result (calibration of the QD was performed without varying the polarization of the beam). Finally, in a typical GH experiment (see below) we have $\chi$/SUM = $\rm 10^{-3} - 10^{-5}$ (with SUM $\sim$ 300-350 mV).

The mirror substrate is made of Duran ceramic glass, has a diameter of 10 cm and a $\lambda$/20 surface flatness. The substrate has been coated by LASEROPTIK \cite{LASEROPTIK} with a chromium film of few nm (to ensure proper sticking of Au) and then a Au film of 200 nm; this thickness is an order of magnitude larger than the penetration depth (skin depth) of the 826-nm radiation in the Au film. The surface roughness of the mirror was determined with a WYKO optical interferometer to be 0.8 nm rms; a scan with an atomic force microscope yielded 1.3 nm rms.

When the polarization is switched from $p$ to $s$, the LCVR introduces a small but noticeable angular tilt of the beam. This tilt was measured by varying the distance of the quadrant detector from the LCVR (without the mirror installed) and was found to be of the order of $3\cdot10^{-7}$ rad. Our actual experimental data, taken with the mirror installed, were corrected for this angular tilt; the correction, when expressed as a contribution to the GH shift, was typically of the order of 50 nm.

\begin{figure}
\centering\includegraphics[scale=0.45]{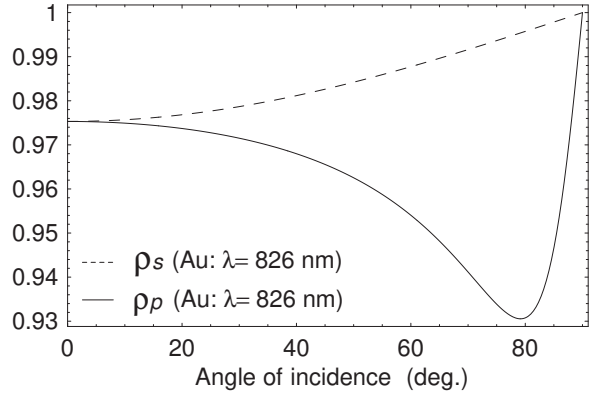}% Here is how to import EPS art
\caption{\label{fig:epsart} Calculated reflectivity of Au at a wavelenght of 826 nm, as a function of the angle of incidence. We verified experimentally that the difference in the reflectivity for the $s$ and $p$ polarized beams is maximal at 80$^{\circ}$.}
\end{figure}

The measurements were performed by reading directly the X channel of the QD with a lock-in amplifier. It is important to note that the reflectivity of Au is different for $s$ and $p$ polarization. Specifically, the measured signal ($\Delta\chi$) is given by $\Delta\chi=C\cdot(\rho_{p} I d_{p}-\rho_{s} I d_{s})$ where $\rho_{p}$ ($\rho_{s}$) is the reflectivity of Au for the $p$ ($s$) polarized beam (Fig. 4) and $ d_{p}$ ($ d_{s}$) the beam centroid position. From this signal it is possible to retrieve the GH shift by simply dividing by $C \rho_{s} I$ (or $C \rho_{p} I$). Noting that $ d_{p}- d_{s}=D_{p}-D_{s}$ it is easy to show that:
\begin{equation}
\frac{\Delta\chi}{C \rho_{s} I}=\frac{(\rho_{p}-\rho_{s})}{\rho_{s}}\cdot d_{p}+ (D_{p}-D_{s}).
\end{equation}
In princple we can easily make the first term in Eq. (4) much smaller than the second one by carefully centering the beam (with the help of the translation stages and the nanovoltmeter). This typically gives $d_{p}< $ 300 nm, moreover, as can be seen from Fig. 4, we have $(\rho_{p}-\rho_{s})/\rho_{s}<$ 0.06 so that the first term in Eq. (4) is at most 18 nm.

Our experimental results are presented in Fig. 5 as solid dots. Data are for the difference between $D_{p}$ and $D_{s}$, that is the quantity that is accessible for our experiments. We varied the angle of incidence from 20$^{\circ}$ to 87$^{\circ}$. We took data every 5$^{\circ}$ in the interval between 20$^{\circ}$ to 80$^{\circ}$, and every 1$^{\circ}$ from 80$^{\circ}$ to 87$^{\circ}$. We noticed a dependence of the GH shift on the beam position on the mirror; therefore, after each series of data we took care to rotate the mirror in order to average these position-dependent effects. Since the lock-in amplifier is a phase sensitive detector it was easy to directly check that the relative positions of the $p$ and $s$ polarized reflected beam are those predicted by theory (see Fig. 1). The lower curve in Fig. 5 is the theoretical prediction for Au ($\hat{n}_{2}=0.188+ i 5.39$). The agreement with our experimental data is very good. The error bars that are reported in the graph represent the standard deviation of the measurements at each angle (we have 10 different measurements per angle). It is evident that the error bars increase with the angle of incidence and decrease again for grazing angles of incidence. The reason for this behavior is unclear. We have also measured the beam displacement in the plane orthogonal to the plane of incidence (Y channel of the QD); in this case there should be no GH shift. Corresponding data (open dots) are reported in the same graph (Fig. 5). We observe on the two channels the same behavior for the error bars.

\begin{figure}
\centering\includegraphics[scale=0.85]{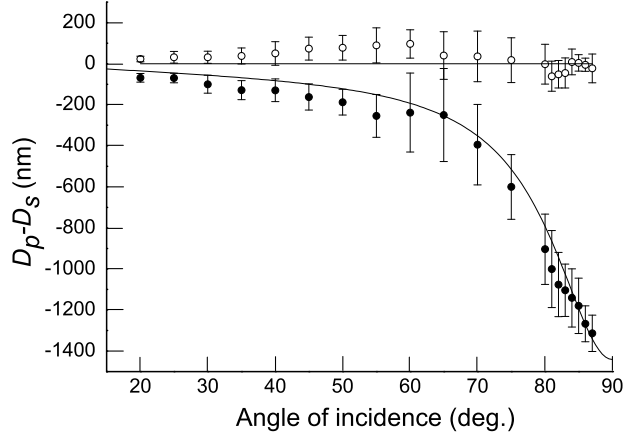}% Here is how to import EPS art
\caption{\label{fig:epsart} Measured Goos-H$\rm \ddot{a}nchen$ shifts, i.e. the difference between $D_{p}$ and $D_{s}$ as a function of the angle of incidence. Experimental data are shown as solid dots and the corresponding theoretical curve has been derived from Fig. 2. The open dots show displacements orthogonal to the plane of incidence; the theoretical line in this case indicates zero displacement.}
\end{figure}

The error bar at 20$^{\circ}$ indicates apparently the resolution (20 nm) of our set up. We find that the resolution is limited both by the LCVR and by unavoidable drifts in the system since they limit the integration time, which was 10 s for our measurements. We are not limited by the electrical noise of the QD which is of the order of $\rm 10^{-7} V_{rms}/ \surd Hz$. Of course, resolution must not be confused with accuracy; the latter refers to the absolute measurements of the position of the beam centroid on the QD. Limited accuracy combined with the difference in reflectivity of Au for $s$ and $p$ polarization (Fig. 4) can generate an error that is angle dependent (Eq. 4). In this case, however, the largest error bar should occur at 80$^{\circ}$, whereas in the experiment we find this at 60$^{\circ}$ - 65$^{\circ}$.

Finally we note that the reflectivity of the metallic surface depends on the angle of incidence, contrary to the usual case of TIR from a dielectric interface far above the critical angle. If one considers the plane wave spectrum of the incident gaussian beam, one can see that some plane wave components will experience lower reflection than some others, leading to an angular shift of the reflected beam \cite{Tamir86, Nasalski89}. We estimate this angular shift for our experiment to be $\rm 1.7\cdot10^{-8}$ rad for $p$ polarization and $\rm 5\cdot10^{-10}$ rad for $s$ polarization (formula 21 of ref \cite{Tamir86}). Since the standard distance from the mirror to the QD was 0.15 m the angular shift is expected to lead to a maximum apparent GH shift of 2.5 nm, i.e. a negligible error. We confirmed this in an experiment in which we varied the distance of the QD from the mirror without any observable effect on the measured GH shift.

\section{Conclusion}

The GH shift for $p$ polarized light in metals has been predicted to be negative; we have confirmed this behavior by measuring the GH shift as a function of the angle of incidence. The large negative shift observed for $p$ polarization indicates the presence of a backward flux of energy at a bare metal surface \cite{Lai00}. Further work is required to understand the intriguing angular dependence of the error bars. Another point of further study concerns the use of other metals than Au, in particular metals which have larger losses than Au. Preliminary calculations show that the existence of a large, negative GH shift (for $p$ polarization) is persistent for relatively large losses, namely up to $\epsilon''\approx |\epsilon'|$, where $\epsilon'=n_{2}^{2}-k^{2}$ and $\epsilon''=2n_{2}k$. (Note that the low-loss condition $\epsilon''\ll |\epsilon'|$ corresponds to $k\gg n_{2}$.)

\
\
\

\textbf{Acknowledgments}

This project is part of the scientific program of FOM. We thanks Dr. G. Gubbels of TNO for performing the WYKO measurements and Dr. N. Matsuki and Dr. C.C.G. Visser of DIMES-Technology Center for performing the AFM measurements. We also thank dr. J. G$\rm \ddot{o}$tte for preliminary calculations of the $\epsilon''$-dependence of the GH shift.


\begin{thebibliography}{99}
\bibitem{Aiello} Note that the expressions for the phase in ref. \cite{Giles82} contain a misprint.
\end{thebibliography}
\end{document}